\documentclass[nofootinbib, pre, twocolumn]{revtex4}
\usepackage{amsmath}
\usepackage{amssymb}
\usepackage{graphicx}
\usepackage{bm}
\usepackage{braket}
\usepackage[version=4]{mhchem}
\usepackage{color, soul}
\usepackage[dvipsnames]{xcolor}
\usepackage{bbm}
\graphicspath{ {./} }
\usepackage{float}
\usepackage{soul}
\usepackage{capt-of}
\usepackage{hyperref}

\DeclareMathOperator{\id}{\mathbbm{1}}
\newcommand{\beq}{\begin{equation}}
\newcommand{\eeq}{\end{equation}}

\begin{document}

\title{Quantum Algorithm for Simulating Molecular Vibrational Excitations}

\author{Soran Jahangiri}
\affiliation{Xanadu, Toronto, ON, M5G 2C8, Canada}
\author{Juan Miguel Arrazola}
\affiliation{Xanadu, Toronto, ON, M5G 2C8, Canada}
\author{Nicol\'as Quesada}
\affiliation{Xanadu, Toronto, ON, M5G 2C8, Canada}
\author{Alain Delgado}
\affiliation{Xanadu, Toronto, ON, M5G 2C8, Canada}

\begin{abstract}

The excitation of vibrational modes in molecules affects the outcome of chemical reactions, for example by providing molecules with sufficient energy to overcome activation barriers. In this work, we introduce a quantum algorithm for simulating molecular vibrational excitations during vibronic transitions. We discuss how a special-purpose quantum computer can be programmed with molecular data to optimize a vibronic process such that desired modes get excited during the transition. We investigate the effect of such excitations on selective bond dissociation in pyrrole and butane during photochemical and mechanochemical vibronic transitions. The results are discussed with respect to experimental observations and classical simulations. We also introduce quantum-inspired classical algorithms for simulating molecular vibrational excitations in special cases where only a limited number of modes are of interest.

\end{abstract}

\maketitle

\section{Introduction}
    
The stability and reactivity of molecules can be influenced by the way they vibrate. Thermal and light-induced vibrational excitations can provide molecules with enough kinetic energy to overcome activation barriers along specific reaction coordinates. Molecular vibrations explain the mechanism of important natural phenomena such as enzyme-catalyzed hydrogen transfer in biological systems \cite{Hay_2012}. They also affect the stability of atmospheric compounds that are subject to sunlight-driven vibrational excitation \cite{Vaida_2014} and provide a means to control the outcome of chemical reactions by selectively exciting vibrational modes that contribute to a desired reaction coordinate \cite{Crim_2008, Chen_2018, Crim_1999, Heyne_2019}. This is important in chemical reactions that are triggered by an abrupt change in the electronic state of a molecule. In such vibronic transitions, the change in the electronic state is usually accompanied by vibrational excitations that can initiate chemical reactions at the new electronic state \cite{Crim_1996}. The ability to engineer the vibrational state of a molecule during a vibronic transition can in principle be used to affect the outcome of chemical reactions \cite{Crim_1996, Epshtein_2011, Grygoryeva_2019}.

A vibronic transition can be mediated by the absorption of light that excites molecules to higher-energy electronic states. The energy provided by the absorbed photons partially transfers into vibrations that can help to overcome reaction barriers of predissociative electronic states \cite{Crim_1996}. A similar mechanism has also been reported for the reactions of molecules on metal surfaces where vibrational excitations are induced by electron transfer from a scanning tunnelling microscopy tip to the adsorbed molecule \cite{Motobayashi_2014}. This charge-transfer process corresponds to a vibronic transition between two different charge states of the molecule. The vibrational excitations initiated by such transitions help to break covalent bonds in the molecule \cite{Maksymovych_2008, Chen_2019, Erpenbeck_2018, Jeong_2017, Stipe_1997}, affect the adsorption of the molecule at the surface \cite{Pascual_2003, Sainoo_2005}, and change its electron transport properties \citep{cuevas2010molecular}. Vibrational excitation in the sudden-force regime of a mechanochemical processes, which can be considered as a vibronic transition to a force-modified potential energy surface, has also been shown to provide excess energy that helps with crossing barriers along a mechanochemical reaction \cite{Rybkin_2017}.

In all of the processes mentioned, the change in the molecular electronic state is accompanied by vibrational excitations that have important effects on the chemical properties of a molecule. However, predicting the probabilities of excitation to all vibrational levels is challenging for transitions that involve simultaneous changes in the vibrational and electronic states of molecules \cite{Jacob_2019}. Furthermore, the time-dependent redistribution of the vibrational energy between the localized modes of a molecule that undergoes a vibronic transition also affects the stability of specific bonds \cite{Sparrow_2018}. Simulation of such vibrational quantum dynamics can also be challenging for conventional quantum chemistry methods.   

Gaussian boson sampling (GBS) \cite{Hamilton_2017} is a platform for photonic quantum computation that has a variety of use cases~\cite{Bradler_2018, Arrazola_2018, Banchi_2019, Jahangiri_2020, Schuld_2020, Bromley_2020}, including the simulation of vibronic spectra of molecules \cite{huh2015boson, quesada2019franck, sawaya2019quantum}. When a GBS device is programmed with the appropriate molecular parameters, the distribution of photons in the optical modes of the device can be used to obtain the distribution of vibrational quanta in the molecule during a vibronic transition. This information can also be obtained from classical algorithms, but their computational complexity increases rapidly with molecular size, rendering such classical methods inefficient for large molecules \cite{Jacob_2019}. This makes GBS a candidate for efficient simulation of vibrational excitation and vibrational quantum dynamics of molecules undergoing vibronic transitions. Simulation of such excitations allows optimizing a vibronic process such that specific target modes become vibrationally excited. The ability to control the final vibrational states during a vibronic process helps, essentially, to affect chemical reactions by selectively activating modes that become reaction coordinates in a desired reaction channel.

In this work, we introduce a quantum algorithm based on GBS for simulating the excitation of vibrational modes during vibronic transitions. We also introduce quantum-inspired classical algorithms that can be employed in special cases where excitations in only a few modes are important. We use these algorithms to explore the vibrational excitations in pyrrole and butane during vibronic transitions. In the case of pyrrole, the transition is initiated by photoexcitation; for butane, it occurs due to a sudden-force mechanochemical excitation. We  discuss the results of the algorithms with respect to the corresponding experimental observations and classical simulations. We also outline procedures for selective excitation of vibrational modes by optimizing external factors that mediate a vibronic process or affect the vibrational quanta distribution during such transitions.

In Sec.~\ref{sec:theory}, we discuss the theory of vibronic transitions and Gaussian boson sampling. We then describe the quantum algorithm in Sec.~\ref{sec: Algorithm} and explain how quantum-inspired classical algorithms can be used when only a small number of modes are of interest. In Sec.~\ref{sec:apps}, we explore applications of the algorithm in the photoexcitation of pyrrole and in force-induced vibrational excitations of butane. Finally, we summarize and discuss the main results of the article in Sec.~\ref{sec:conc}.

\section{Theory}\label{sec:theory}

The theoretical concepts are explained in this section. We begin with a short description of vibronic transitions, introducing the central concepts of Franck-Condon factors, Duschinsky transformations, and Doktorov operators in Sec.~\ref{sec:vib}. We also describe the computational challenges involved in simulating vibronic transitions. In Sec.~\ref{sec: GBS}, we give an overview of Gaussian boson sampling and explain how these photonic devices can be used to simulate vibronic transitions. In Sec.~\ref{sec:prog}, we discuss how a GBS device can be programmed with molecular data. We conclude by discussing the simulation of vibrational excitations when the vibrational levels of a molecule are pre-excited using infrared light in Sec.~\ref{sec:init}.

\subsection{Vibronic transitions}\label{sec:vib}

A vibronic transition involves simultaneous changes in the electronic and vibrational states of a molecule. The Franck-Condon (FC) approximation \cite{sharp1964franck, Barone_2009} states that the probability of a vibronic transition is determined by FC factors, which are the squares of the overlap integrals between the vibrational wave functions of the initial and final states. Predicting the probabilities of vibrational excitations during a vibronic transition requires estimating the FC factors for all of the possible transitions between the vibrational states of the two electronic states. The total number of FC factors that need to be estimated depends on the number of vibrational modes ($M$) and the maximum number of vibrational quanta ($K$) in all modes as~\cite{ezSpectrum}:
\beq \label{Eq: nFC}
n_{FC} = \left(1 + \sum_{k=1}^{K} \frac{(M + k -1)!}{(M - 1)!k!} \right)^2.
\eeq 
Under the harmonic approximation, the FC factors can be written as integrals over the eigenstates of an $M$-dimensional harmonic oscillator. When the normal coordinates of the initial and final electronic states are the same, these multidimensional integrals are products of multiple one-dimensional integrals, which can be computed analytically.

More generally, normal coordinates are mixed during a vibronic transition, which makes computing FC factors more challenging. Mathematically, the normal coordinates of the initial and final states, $\bm{q}$ and $\bm{q}'$, are related to each other via the Duschinsky transformation \cite{Duschinsky_1937}:
\beq
\bm{q'} = U_D\bm{q} + \bm{d},
\eeq
where $U_D$ is the Duschinsky matrix, an orthogonal and in general non-diagonal matrix that is related to the overlap between the normal modes, and $\bm{d}$ is a real vector that describes the change in the molecular geometries of the initial and final states. The Duschinsky transformation prevents the FC integrals to be reduced to products of one-dimensional integrals. In this case, the integrals can be computed recursively starting from the overlap integral between the ground vibrational states of the initial and final electronic states involved in the vibronic transition \cite{Doktorov_1977, Berger_1998, Koziol_2009, ezSpectrum}. The computational expenses associated with such recursive approach and the large number of integrals (given in Eq.~\eqref{Eq: nFC}) make the complexity of evaluating the FC profile, and vibrational excitations,
to scale combinatorially with the size of the molecule \cite{huh2015boson}. Quantum algorithms could therefore be helpful in enabling a more efficient simulation of vibronic transitions that circumvents the need to compute these integrals.

\begin{figure*}
\includegraphics[width=1.5 \columnwidth]{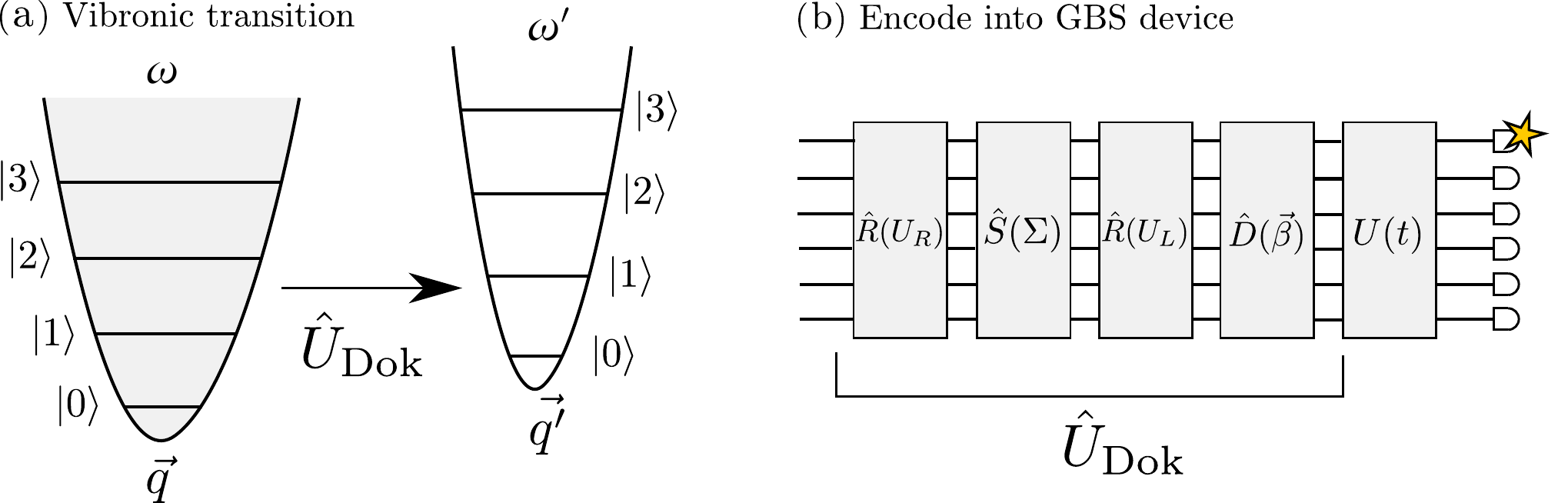}
\centering
\caption{(a) Potential energy curve of a diatomic molecule undergoing a vibronic transition. The potential is assumed to be harmonic for both the initial and final states. A vibronic transition can be represented in terms of a Doktorov transformation $\hat{U}_{\text{Dok}}$, which is determined from the normal-mode frequencies $\omega, \omega'$ and normal coordinates $\bm{q}, \bm{q'}$. (b) The Doktorov operator can be decomposed in terms of displacement $\hat{D}(\bm{\beta})$, squeezing $\hat{S}(\Sigma)$, and rotation $\hat{R}(U_L)$, $\hat{R}(U_R)$ operations. These can be implemented in a GBS device to prepare the final state after the transition. A time-dependent transformation $U(t)$ can also be implemented to simulate the vibrational quantum dynamics. The excitations are sampled by measurements in the photon-number basis.}\label{fig:algorithm}
\end{figure*}

The FC factors can also be written in terms of the initial and final vibrational Fock states \cite{huh2015boson}:
\beq \label{Eq: FC_Dok}
    F(\bm{m},\bm{n}) = \left | \bra{\bm{m}}\hat{U}_{\text{Dok}}\ket{\bm{n}} \right | ^ 2.
\eeq
Here $\ket{\bm{n}}=\ket{n_1, n_2,\ldots, n_M}$ is a state with $n_i$ vibrational quanta in the $i^\text{th}$ normal mode of the ground electronic state,  
and $\ket{\bm{m}}=\ket{m_1, m_2,\ldots, m_M}$ is a state with $m_j$ vibrational quanta in the $j^\text{th}$ normal mode of the excited electronic state. Finally, $\hat{U}_{\text{Dok}}$ is the Doktorov operator \cite{huh2015boson} which can be decomposed in terms of multi-mode displacement $\hat{D}(\beta)$, squeezing $\hat{S}(\Sigma)$, and generalized rotation $\hat{R}(U_L)$, $\hat{R}(U_R)$ operators as~\cite{quesada2019franck}:
\beq\label{Eq: Dok_Gaussian}
    \hat{U}_{\text{Dok}} = \hat{D}(\bm{\beta})\hat{R}(U_L) \hat{S}(\Sigma) \hat{R}(U_R),
\eeq
where $U_L$ and $U_R$ are unitary matrices, $\Sigma$ is a diagonal matrix, and $\bm{\beta}$ is a vector of displacements.

The squeezing and displacement operators are related to the changes in the vibrational frequency and the equilibrium distance of the molecule during the transition, respectively. Similarly, the rotation operators correspond to the rotation of the normal modes of the initial electronic state in the normal mode basis of the final state. In this framework, the relation between the normal coordinates of the initial and final states and the corresponding transformation of the bosonic operators for the vibronic transition are contained in $\hat{U}_{\text{Dok}}$. The states $\ket{\bm{n}}$ and $\ket{\bm{m}}$ only contain information about the number of vibrational quanta in the initial and final states, while the FC factor determines the probability of observing an excitation to the state $\ket{\bm{m}}$ starting from the state $\ket{\bm{n}}$.
For any given molecule, $\hat{U}_{\text{Dok}}$ can be obtained from the vibrational normal modes and frequencies and the equilibrium geometries of the initial and final electronic states. The vibronic transition of a diatomic molecule, represented by a single-mode harmonic oscillator, is shown in Fig.\ref{fig:algorithm}(a) in terms of a Doktorov transformation.

\subsection{Gaussian boson sampling}\label{sec: GBS}

Gaussian Boson Sampling (GBS) is a platform for photonic quantum computation. A GBS device consists of a multimode linear-optical framework in which squeezed light is injected into each mode, passed through a linear-optical circuit and finally measured at the output. The device can be set up based on the parameters of the Doktorov operator for a given molecule according to the scheme presented in Fig.\ref{fig:algorithm}(b). This device can be programmed to compute FC profiles by exploiting the equivalence between photons in optical modes and vibrational quanta in the normal modes of the molecule~\cite{huh2015boson}. This correspondence also allows programming a GBS device to simulate the distribution of vibrational quanta during a vibronic transition in order to determine the excitation of specific vibrational modes of the final electronic state.

The probability of observing an output state $\ket{\bm{m}}=\ket{m_1, m_2, \ldots, m_M}$ in a GBS device programmed with $\hat{U}_{\text{Dok}}$ is:
\beq \label{Eq: GBS_Prob0}
    \Pr(\bm{m},\bm{n}) = \left | \bra{\bm{m}}\hat{U}_{\text{Dok}}\ket{\bm{n}} \right | ^ 2.
\eeq
where $\ket{\bm{n}}=\ket{n_1, n_2, \ldots, n_M}$ refers to the initial state. The similarity between Eqs.~\eqref{Eq: GBS_Prob0} and~\eqref{Eq: FC_Dok} makes it possible to encode the chemical information characterizing a vibronic transition into a GBS distribution, then sample from it to determine the statistics of the resulting vibrational excitations.

While generating samples from a real GBS device is relatively fast, simulating the sampling process is not efficient on classical computers. The probability of observing an output state in a GBS setting is given by~\cite{quesada2019franck}:
\begin{align}\label{Eq: GBS_Prob1}
\Pr(\bm{m},\bm{n})  =  \frac{\text{lhaf}(\mathcal{A}'_{\bm{m},\bm{n}})}{\mathcal{N}}.
\end{align}
In this equation, $\text{lhaf}(\cdot)$ is a matrix function called the loop hafnian \cite{bjorklund2018faster}, $\mathcal{A}'_{\bm{m},\bm{n}}$ is a matrix determined by the covariance matrix and vector of displacements of the Gaussian state prepared by the GBS device and by the distribution of initial and final vibrational quanta $(\bm{m},\bm{n})$. Finally, $\mathcal{N}$ is a normalization constant. These quantities are defined in Appendix \ref{app:gbs_prob}. The complexity of the best-known classical algorithms for sampling from the probability distribution in Eq.~\eqref{Eq: GBS_Prob1} scales exponentially with the total number of photons and polynomially with the number of modes~\cite{quesada2020exact}. For large systems, this makes classical simulation intractable. 

\subsection{Programming a GBS device with molecular data}\label{sec:prog}

Programming a GBS device for a given molecule requires determining the Doktorov operator for that molecule and mapping its parameters to the GBS device. The Doktorov operator is obtained from the Duschinsky matrix and displacement vector. For a given molecule, the Duschinsky matrix is obtained from the eigenvectors of the initial and final state Hessian matrices, $\bm{L}$ and $\bm{L}'$, respectively \cite{Reimers_2001}:
\beq
U_D = (\bm{L}')^T \bm{L}.
\eeq
The displacement vector $\bm{d}$ is related to the Cartesian geometry vectors of the initial and final states, $\bm{x}$ and $\bm{x}'$, as \cite{Reimers_2001}:
\beq
\bm{d} = (\bm{L}')^T m^{1/2} (\bm{x} - \bm{x}'), 
\eeq 
where $m$ is a diagonal matrix containing atomic masses. The quantities $\bm{L}$, $\bm{L}'$ and $\bm{x}$ can be obtained from electronic structure calculations. 

The matrices $U_L$, $U_R$, and $\Sigma$ in Eq.~\eqref{Eq: Dok_Gaussian} are then obtained from the singular value decomposition $J = U_L\Sigma U_R$ of the matrix $J:=\Omega' U_D\Omega^{-1}$. The diagonal matrices $\Omega$ and $\Omega'$ are respectively obtained from the ground and excited state frequencies:
\begin{align}
\Omega &= \text{diag} (\sqrt{\omega_1},...,\sqrt{\omega_M}),\\
\Omega' &= \text{diag} (\sqrt{\omega_1'},...,\sqrt{\omega_M'}).
\end{align} 
Finally, the displacement vector $\bm{\beta}$ in Eq.~\eqref{Eq: Dok_Gaussian} is given by $
\bm{\beta}=\hbar^{-1/2}\Omega'\bm{d}/\sqrt{2}$ where $\hbar$ is the reduced Planck constant.

Once the Doktorov operator has been determined for a molecule, the quantum algorithm can be used to obtain the excitation of vibrational modes during a vibronic transition. Furthermore, as shown in Ref.~\cite{Sparrow_2018}, a photonic quantum device can also be configured to implement the unitary transformation
\beq \label{Eq: Ut}
U(t) = \hat{R}(U_l) e^{-i \hat{H} t/\hbar},
\eeq
to simulate the vibrational quantum dynamics of molecules. In Eq.~\eqref{Eq: Ut}, $U_l$ is a unitary matrix that converts the molecular normal modes to a set of spatially-localized vibrational modes, $\hat{R}(U_l)$ represents an interferometer configured with respect to $U_l$, $\hat{H}$ is the Hamiltonian corresponding to the harmonic normal modes, and $t$ is time.

\subsection{Initial state preparation}\label{sec:init}

The probability of observing a specific transition between two vibrational states during a vibronic process depends on the FC factor that characterizes the transition and also on the population of the initial vibrational states of the molecule. At finite temperatures, the population of the initial vibrational states follows a Boltzmann distribution. The corresponding vibronic transitions can be simulated using a modified GBS setup as discussed in Ref.~\cite{Huh_2017}. The initial vibrational state in a vibronic transition can also be prepared via pre-excitation of specific vibrational levels with infrared (IR) light~\cite{Crim_1996}. The vibronic  photoexcitation from such pre-excited states to higher-energy electronic states can be only simulated by a GBS device if the initial pre-excited state is properly defined. In this section, we show that a vibronic process initiated from such pre-excited states can be simulated with a GBS device when the initial state is a coherent state. 

We start by writing the standard light-matter Hamiltonian in the dipole approximation:
\begin{align}
\hat{H}_{\text{LM}} = \hat{\bm{\mu}} \cdot \bm{E}(t),
\end{align}
where $\bm{\mu}$ is the dipole moment of the molecule that we split in terms of its electronic and nuclear parts as:
\begin{align}\label{Eq: Mu}
\hat{\bm{\mu}} = \hat{\bm{\mu}}_e + \hat{\bm{\mu}}_n = \hat{\bm{\mu}}_e + \sum_i q_i \hat{\bm{r}}_i.
\end{align}
In Eq.~\eqref{Eq: Mu}, $q_i$ is the charge of the $i^\text{th}$ nuclei and $\bm{r}_i$ is its position.
We consider a classical electric field $\bm{E}(t)$ with amplitude $\bm{E}_0$ and oscillating frequency $\omega_0$,
\begin{align}
\bm{E}(t) = \bm{E}_0 e^{-i \omega_0 t} +\bm{E}_0^* e^{i \omega_0 t}.
\end{align}
We assume that the frequency of the IR light, used to excite the initial vibrational state, is far away from any electronic resonance at this stage and thus ignore any contribution due to the electronic dipole moment $\hat{\bm{\mu}}_e$. Furthermore, in the harmonic approximation, we can expand the position of any atom in terms of the centre of mass and the normal coordinates as:
\begin{align}
\hat{\bm{r}}_i = \sum_{j} \bm{c}_{ij} \left(\hat{a}_j^\dagger + \hat{a}_j \right),
\end{align}
where $\bm{c}_{ij}$ are the expansion coefficients of the position of the $i^{th}$ atom in terms of the corresponding normal coordinate represented by its creation and destruction operators. It is also assumed that the centre of mass, if confined, oscillates at a frequency far from $ \omega_0$, so its contribution is neglected. We can now take the normal mode expansion in the last equation and write the light-matter Hamiltonian in the interaction picture (replacing $\hat{a}_j^\dagger \to \hat{a}_j^\dagger e^{i \omega_j t}$) as \cite{sakurai2017modern}:
\begin{align}
H_{\text{LM}}^{I}(t) =& \sum_i q_i \hat{C}_i \cdot \left( \bm{E}_0 e^{-i \omega_0 t} +\bm{E}_0^* e^{i \omega_0 t} \right),
\end{align}
where
\begin{align}
\hat{C}_i = \sum_{j} \bm{c}_{ij} \left(\hat{a}_j^\dagger  e^{i \omega_j t} + \hat{a}_j  e^{-i \omega_j t} \right).
\end{align}
The time-evolution operator associated with this Hamiltonian is \cite{sakurai2017modern}:
\begin{align}
\hat{\mathcal{U}}(t_0,t_1) = \hat{\mathcal{T}}\exp\left( - \frac{i}{\hbar} \int_{t_0}^{t_1} dt H_{\text{LM}}^{I}(t)  \right),
\end{align}
where $\hat{\mathcal{T}}$ is the time-ordering operator. Assuming that the frequency of the electric field is resonant with normal mode $k$, we show in Appendix \ref{app:normal_modes} that the time-evolution operator is simply a displacement in mode $k$:
\begin{align}\label{Eq:time-evolution}
\hat{\mathcal{U}}(t_0,t_1) = \hat{D}\left(- \frac{i}{\hbar} \sum_i q_i  \bm{c}_{ik}\cdot \bm{E_0} \  \{t_1-t_0\} \right),
\end{align}
where $\hat{D}_k(\beta_k) = \exp(\beta_k \hat{a}_k^\dagger - \beta^*_k \hat{a}_k)$ is a displacement in mode $k$ by an amount $\beta_k = - \frac{i}{\hbar} \sum_i q_i  \bm{c}_{ik}\cdot \bm{E_0} (t_1-t_0)$. When this operator is applied to a state with zero vibrational quanta, the result is a coherent state in mode $k$, which is a Gaussian state \cite{gerry2005introductory}. In the Fock basis, a coherent state with parameter $\beta$ can be represented as:
\beq
\ket{\beta_k} = \hat{D}_k(\beta_k)\ket{0} = e^{-\frac{|\beta_k|^2}{2}}\sum_{n=0}^{\infty} \frac{\beta_k^n}{\sqrt{n!}} \ket{n}.
\eeq   

\begin{figure}
\includegraphics[width=0.95 \columnwidth]{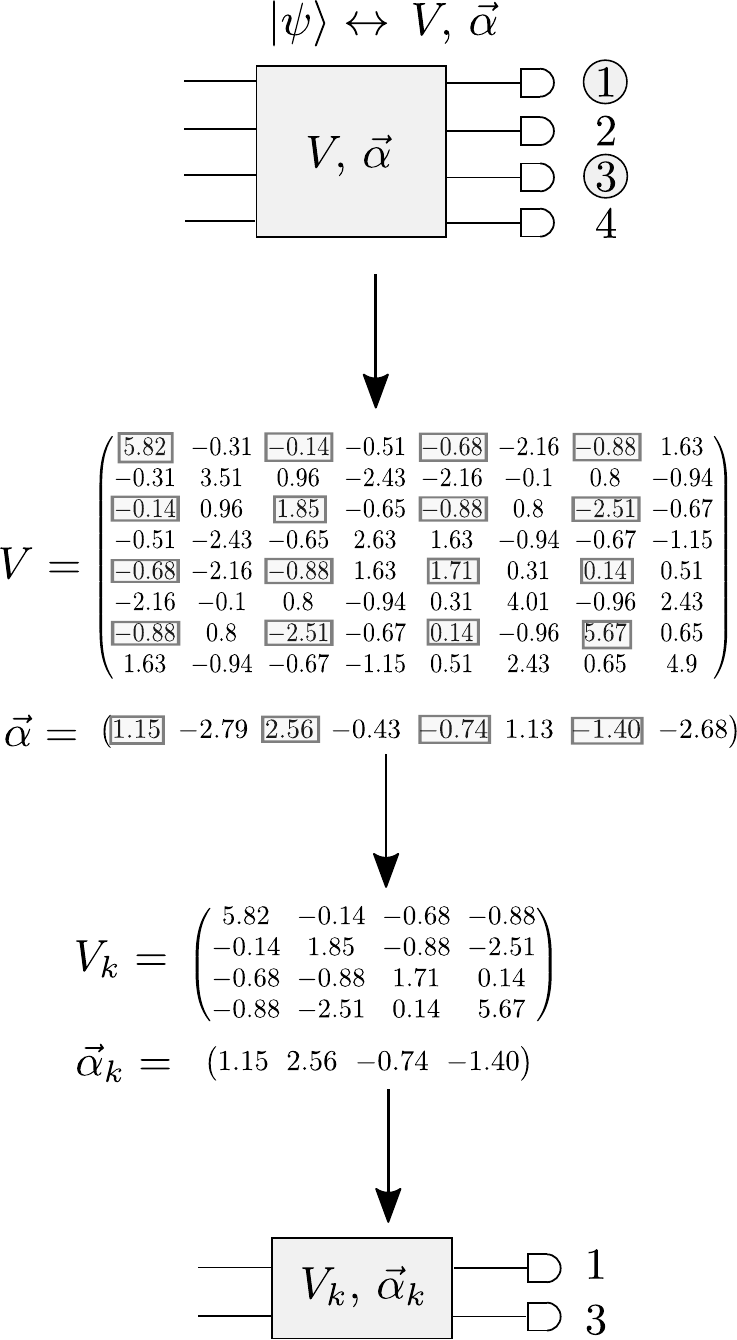}
\centering
\caption{Algorithm for sampling from marginal distributions. The final state $\ket{\psi}$ of the vibrational modes is Gaussian, so it can be represented in terms of a covariance matrix $V$ and a vector of means $\bm{\alpha}$. In this example, the goal is to sample from the marginal distribution of modes 1 and 3 of a four-mode system. The reduced covariance matrix $V_k$ and reduced vector of means $\bm{\alpha}_k$ are obtained by keeping only the entries of columns and rows numbered 1, 3, (1+4)=5, and (3+4)=7. These can then be encoded into a two-mode GBS device to sample from the desired marginal distribution.}\label{fig:marginal}
\end{figure}

\section{Algorithm}\label{sec: Algorithm}

We now outline an algorithm for simulating molecular vibrational excitations during a vibronic transition. The algorithm includes the following steps:

\begin{enumerate}
\item Compute the GBS parameters $U_L$, $\Sigma$, $U_R$ and $\bm{\beta}$ from the input chemical parameters $\Omega$, $\Omega'$, $U_D$ and $\bm{d}$.
\item Use the GBS device to prepare the Gaussian state $\ket{\psi}=\hat{U}_{\text{Dok}}\ket{\psi_i}$ with covariance matrix $V$ and vector of means $\bm{\alpha}$, where $\ket{\psi_i}$ is an initial Gaussian state. 
\item Implement the transformation $U(t)$ to simulate the vibrational quantum dynamics in the localized modes.  
\item Generate samples of the form $\ket{\bm{m}}=\ket{m_1, m_2,\ldots, m_M}$ by measuring the output state in the photon-number basis.
\item Repeat these steps sufficiently many times to obtain the desired statistics about the distribution of vibrational excitations.
\end{enumerate}

In certain situations, only a few of the vibrational modes are of interest and it suffices to sample from their marginal distribution. Gaussian states are uniquely specified by their covariance matrix $V$ and vector of means $\bm{\alpha}$~\cite{picinbono1996second}, so computing marginal distributions is straightforward; the reduced states can be readily obtained from the covariance matrix and vector of means.

For simplicity and without loss of generality, consider the marginal distribution of the first $k$ modes of an initial state with $M$ modes. The $2k\times 2k$ reduced covariance matrix $V_k$ is obtained by selecting the $(i, i+M)$ rows and columns of the original covariance matrix $V$, for $i=1,2,\ldots, k$. Similarly, the reduced vector of means $\bm{\alpha}_k$ is constructed by keeping only the first $k$ entries of the original vector $\bm{\alpha}$. The marginal distribution of the first $k$ modes is then also given by Eq.~\eqref{Eq: GBS_Prob1}, with the exception that all quantities are defined with respect to $V_k$ and $\bm{\alpha}_k$. This process is illustrated in Fig.~\ref{fig:marginal}.

When $k$ is sufficiently small, it is possible to employ existing classical algorithms \cite{quesada2020exact} to simulate the resulting $k$-mode GBS device, thus leading to a quantum-inspired method for simulating vibrational excitations in molecules.

\section{Applications}\label{sec:apps}
In this section, we apply the GBS algorithm to simulate the vibrational excitations in pyrrole and butane during vibronic processes mediated by photoexcitation and mechanochemical excitation. In both applications, we explore situations that require generating GBS samples and also investigate cases where implementing the quantum-inspired algorithm is sufficient. These examples are used to showcase the ability of the quantum algorithm in determining the effect of selective vibrational pre-excitation and the magnitude of external mechanical force on the distribution of vibrational quanta in molecules during vibronic transitions. We explain the results of the quantum algorithm with respect to those of experimental investigations and classical simulations. The calculations performed here are based on the harmonic approximation. Anharmonic effects can also be included in the quantum algorithm, as discussed in Refs.~\cite{sawaya2019quantum, McArdle_2019}. We also perform the quantum simulations at zero temperature to make the calculations computationally affordable.

The electronic structure of the ground and excited states of pyrrole were computed, respectively, using the Coupled-Cluster method at the level of singles and doubles excitations (CCSD) \cite{Piecuch_2002} and its extension to model the excited states, the equation-of-motion CCSD (EOM-CCSD) \cite{Piecuch_2002, Kowalski_2004, Wloch_2005}. The Pople basis set 6-31+G(d) \cite{Ditchfield_1971} was used throughout. Single-point energy calculations with Dunning's correlation-consistent basis set augmented with diffuse functions, aug-cc-pVDZ \cite{Dunning_1989}, were performed to evaluate the accuracy of the smaller basis set in predicting excitation energies. The calculations for butane were performed with density functional theory \cite{Hohenberg_1964} using the hybrid density functional B3LYP \cite{Lee_1988,Becke_1993,Stephens_1994} and aug-cc-pVDZ basis set. All electronic structure calculations were performed with the general atomic and molecular electronic structure system (GAMESS) \cite{Schmidt_1993,Gordon_2005}. The sampling algorithm was implemented by simulating GBS devices using Strawberry Fields~\cite{killoran2019strawberry} and The Walrus~\cite{gupt2019walrus}.

\subsection{Photoexcitation of pyrrole}

The photochemistry of pyrrole has been the subject of several experimental and theoretical investigations \cite{Vallet_2005,Lan_2007, Epshtein_2011,Wu_2015, Grygoryeva_2019, Ashfold_2006}. The first excited singlet electronic state of pyrrole has an experimental excitation energy of 5.22 eV \cite{Flicker_1976}. The excitation energy computed with EOM-CCSD/aug-cc-pVDZ performed on the geometry optimized with EOM-CCSD/6-31+G(d) is 5.22 eV, in perfect agreement with the experimental value. This indicates that the excited state geometry of pyrrole obtained using the smaller basis set 6-31+G(d) is a good approximation for the purposes of our simulations.

\begin{figure}[t]
\includegraphics[scale=0.62]{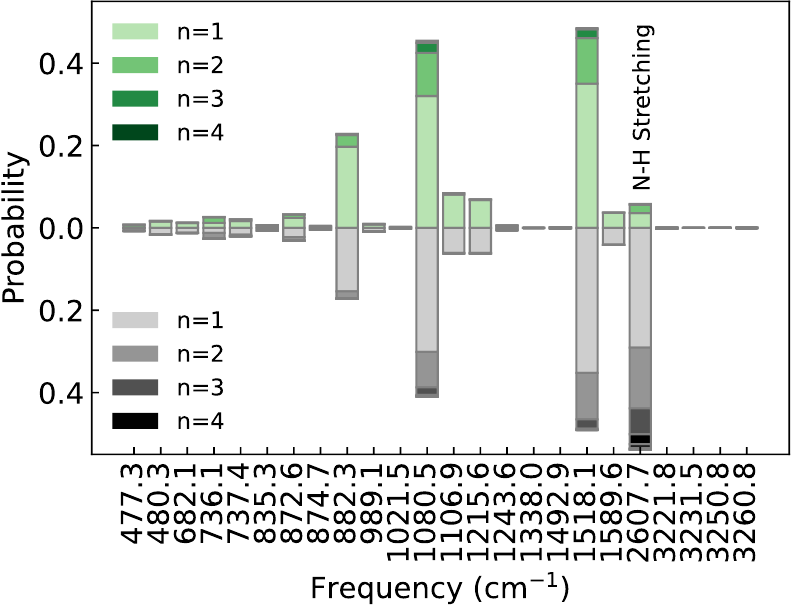}
\centering
\caption{Single-mode marginal distributions of pyrrole during a vibronic transition from ground to the first excited state without (top) and with (bottom) pre-excitation of the ground state N-H stretching mode. The mode with the vibrational frequency of 2607.7 cm$^{-1}$ corresponds to the stretching of the N-H bond in the final electronic state. The number of vibrational quanta in each mode is represented by $n$.
}\label{fig:pymarg}
\end{figure}

\begin{figure}[b]
\includegraphics[scale=0.2]{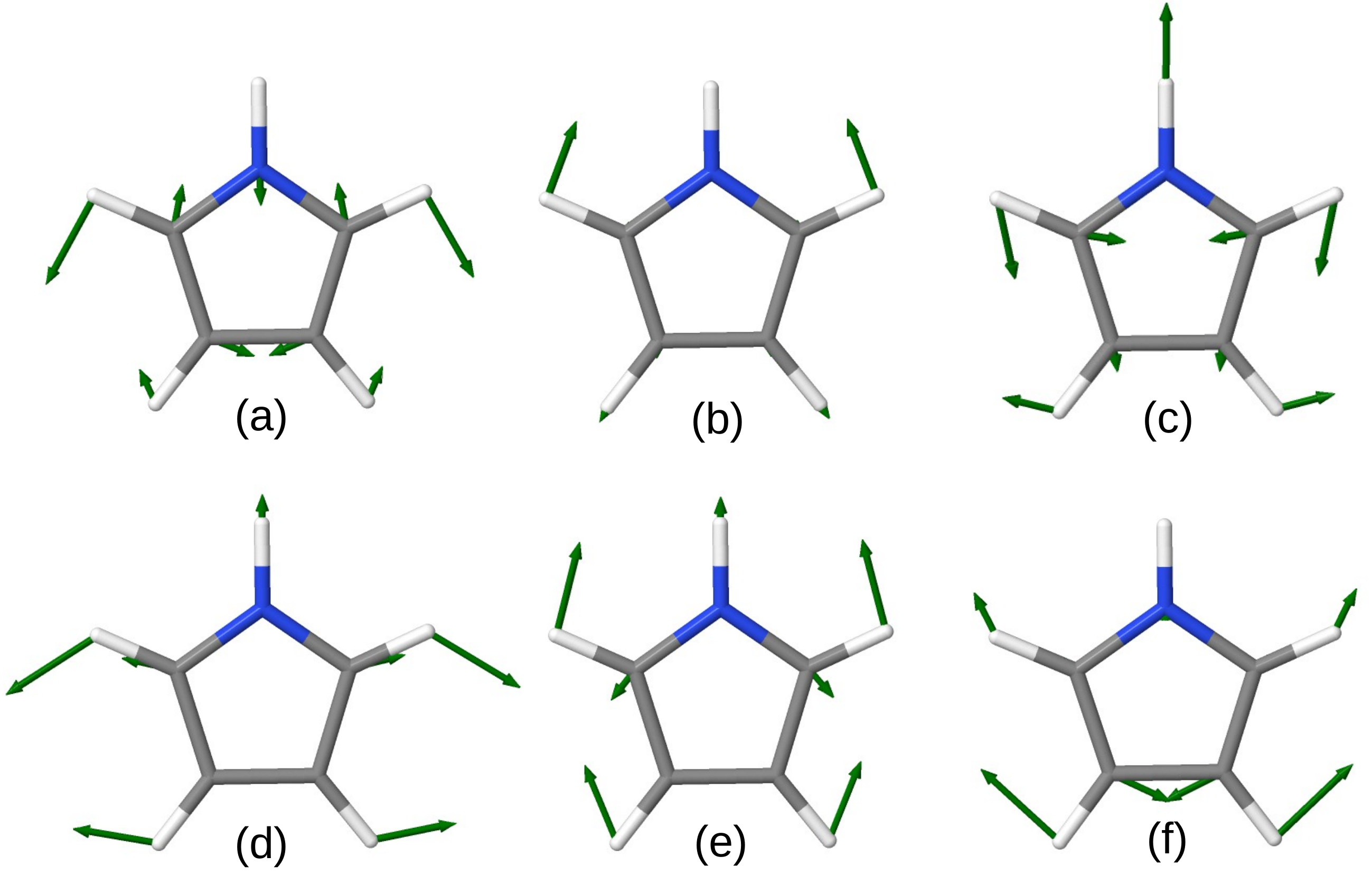}
\centering
\caption{Vibrational normal modes of pyrrole with frequencies of (a) 1518.1 cm$^{-1}$, (b) 1080.5 cm$^{-1}$ and (c) 882.3 cm$^{-1}$, which become highly excited during the vibronic transition. Two modes at the excited electronic state that include stretching of the C-N bonds, with vibrational frequencies of (d) 1215.7 cm$^{-1}$ and (e) 1589.7 cm$^{-1}$ are excited simultaneously as a result of the pre-excitation of the ground electronic state mode with a frequency of (f) 1462.2 cm$^{-1}$.}\label{fig:modes}
\end{figure}

We employ the GBS algorithm to determine the distribution of vibrational excitations of pyrrole after a vibronic transition from the electronic ground state to the first electronic excited state. We first simulate the vibrational excitation of the normal modes of pyrrole at the excited electronic state with and without the vibrational pre-excitation of the nitrogen-hydrogen (N-H) stretching mode at the ground electronic state. We then investigate the effect of vibrational pre-excitation of the ground electronic state normal modes on the simultaneous excitation of two carbon-nitrogen (C-N) stretching modes at the excited electronic state.

The marginal distributions of the normal modes of pyrrole, which determine the probability of vibrational excitations in a single mode, are plotted in Fig.~\ref{fig:pymarg}. These distributions were calculated from the GBS samples obtained for the vibronic transition without the vibrational pre-excitation of the N-H stretching mode. The distributions in Fig.~\ref{fig:pymarg} demonstrate that the normal modes with frequencies 1518.1 cm$^{-1}$, 1080.5 cm$^{-1}$, and 882.3 cm$^{-1}$ become highly excited during the vibronic transition. These vibrational modes are illustrated in Fig.~\ref{fig:modes}. However, the vibrational normal mode that corresponds to the stretching of the N-H bond in the final electronic state, with a vibrational frequency of 2607.7 cm$^{-1}$, is not significantly excited during the vibronic transition (see Fig.~\ref{fig:pymarg}).

Experimental investigations show that excitation of the N-H stretching mode at the electronic ground state, before the initiation of the vibronic transition, enhances its dissociation \cite{Grygoryeva_2019}. This experimental observation can be explained by assuming that the vibrational energy in the N-H stretching mode of the ground state is preserved during the vibronic transition. To validate this assumption, we compute marginal distributions for the vibronic transition in which the N-H stretching mode of the ground state is initially excited. The results, which are shown in Fig.~\ref{fig:pymarg}, demonstrate that pre-excitation of the mode leads to significant vibrational excitation of the corresponding N-H mode after the transition. Since the N-H dissociation energy barrier is low at the excited state (6.0 kcal mol$^{-1}$ \cite{Vallet_2005}), this in principle increases the rate of N-H dissociation, as observed in the experimental investigation \cite{Grygoryeva_2019}. It is noted that a quantitative description of the dissociation rate requires more accurate electronic structure calculations and also the inclusion of anharmonic effects.

\begin{figure}[t]
\includegraphics[width=1 \columnwidth]{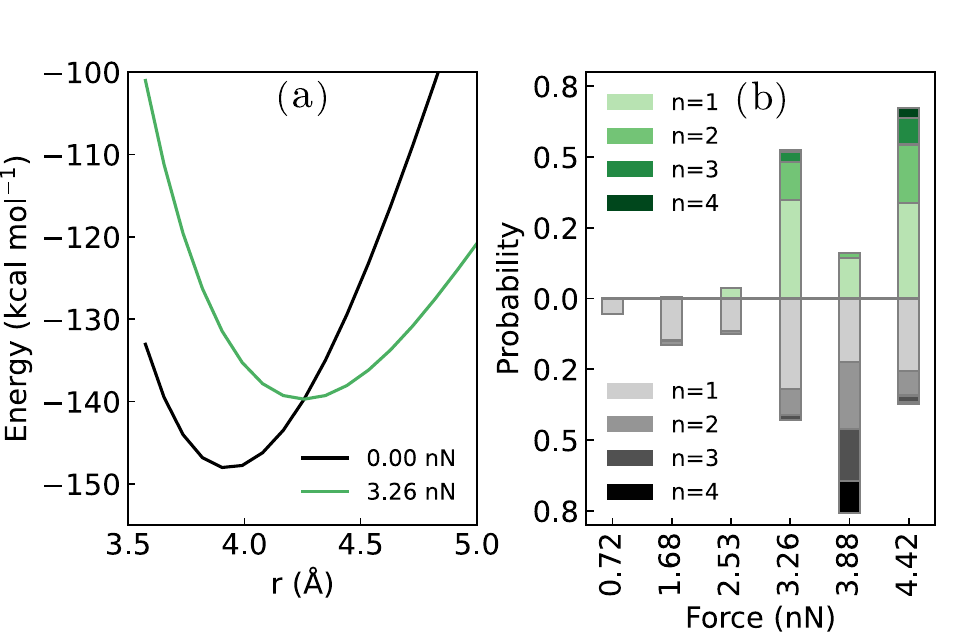}
\centering
\caption{(a) Potential energy curve for stretching the terminal carbon atoms of butane in the presence of an external force. In the sudden-force regime, the application of the external force is abrupt and the molecule is vertically transferred to a new force-modified potential energy surface. This transition to the new potential energy surface is associated with vibrational excitation of the molecule. The distance between the terminal carbon atoms is denoted by $r$. (b) Single-mode marginal distributions of butane during a vibronic transition from ground to the force-modified potential energy surface. The marginal distributions correspond to the inner C-C bond (top) and  an outer C-C bond (bottom) of butane. The number of vibrational quanta in each mode is represented by $n$.}\label{fig:buscanmarg}
\end{figure}

Marginal distributions are valuable when the probability of vibrational excitations in a single mode is of interest. In such cases, the quantum-inspired classical algorithms introduced in Sec.~\ref{sec:prog} can be implemented. However, sampling from the complete probability distribution is more informative when preparation of co-excited vibrational modes in the excited electronic state is needed. We now look at the effect of vibrational pre-excitation of pyrrole modes, at the ground electronic state, on the excitation of two modes at the excited electronic state that include stretching of the C-N bonds. These two modes have vibrational frequencies of 1215.7 cm$^{-1}$ and 1589.7 cm$^{-1}$ and are illustrated in Fig.~\ref{fig:modes}. The choice of these modes is motivated by the ring opening reaction that involves dissociation of both C-N bonds simultaneously. The sampling results demonstrate that pre-excitation of the mode with the vibrational frequency of 1462.2 cm$^{-1}$, illustrated in Fig.~\ref{fig:modes}, leads to the simultaneous excitation of the two normal modes of the excited electronic state that involve C-N stretching. Pre-excitation of this mode, by inserting an average number of one photon, increases the probability of simultaneous excitation of the two C-N stretching modes from 0.2 \% to 7.1 \%. Pre-excitation with a larger number of photons further increases this probability.

\subsection{Mechanochemistry of butane}

\begin{figure}[b]
\includegraphics[scale=0.23]{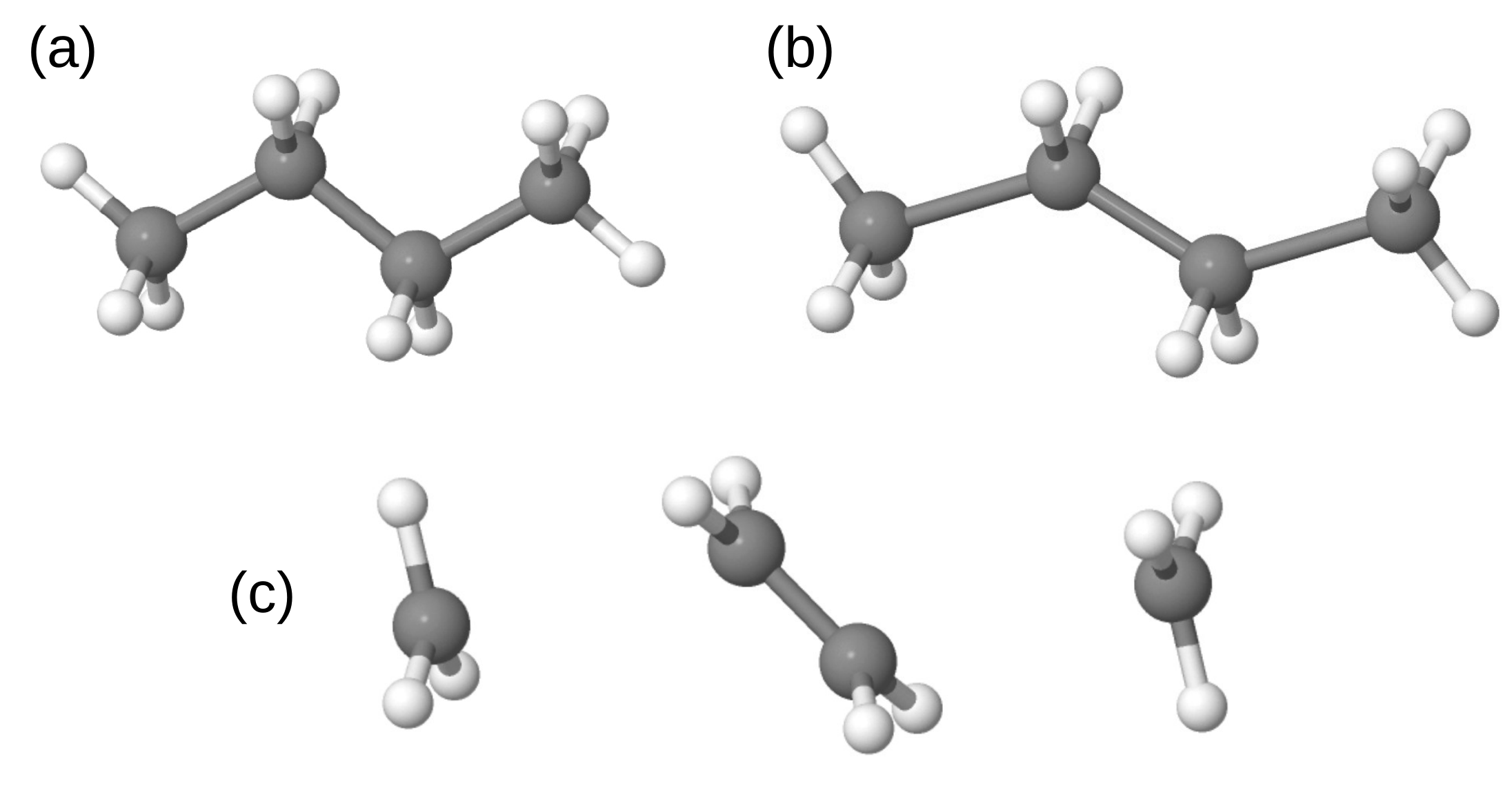}
\centering
\caption{Optimized structure of butane when (a) no stretching force is applied to the terminal carbon atoms, (b) the stretching force reaches the critical value, and (c) the stretching force dissociates the molecule to ethylene and two methyl radicals.}\label{fig:bu}
\end{figure}

In mechanochemistry, a force is applied to a molecule to perform chemical reactions. The mechanism and kinetics of a mechanochemical process might be considerably affected by the rate at which the external force is applied \cite{Smalo_2014}. In the fixed-force regime, the external force acts gradually and the molecular geometry is allowed to relax during each stage of the process. In the sudden-force regime, the external force is applied instantaneously and the molecule is transformed to the force-modified potential energy surface (FM-PES) abruptly. This makes the process of sudden-force mechanochemistry analogous to the Franck-Condon transition in vibronic spectroscopy \cite{Rybkin_2017}. The external energy applied in the sudden-force regime can excite the molecule to higher vibrational energy levels at the FM-PES.

Butane has been investigated in both fixed and sudden-force regimes as a model system for understanding the effect of molecular dynamics on the mechanochemistry of linear alkane chains \cite{Smalo_2014}. In the case of butane, it has been shown that the gradual application of the external field leads to the dissociation of the outer carbon-carbon (C-C) bonds, while in the sudden-force regime, both outer and inter C-C bonds dissociate \cite{Smalo_2014}. The change in the mechanism can be explained by the vibrational excitation of the molecule due to the abrupt application of the external force \cite{Rybkin_2017}. Here we simulate the vibrational excitation of butane on different FM-PES corresponding to different magnitudes of external force applied suddenly to the terminal carbon atoms of the molecule.

We computed the potential energy curve of the molecule as a function of the distance between the terminal carbon atoms (see Fig.~\ref{fig:buscanmarg}) and differentiated the curve numerically to obtain the internal force associated with each distance. The effect of the applied external force on the PES was simulated by constraining the distance between the terminal carbon atoms of butane at a fixed value, corresponding to the desired magnitude of the external force, and optimizing the geometry of the molecule under this constraint. These constrained geometry optimizations, with the fixed distance between the carbon atoms, were followed by vibrational frequency calculations to obtain the normal modes and vibrational frequencies of the FM-PES, analogous to the procedure in Ref. \cite{Rybkin_2017}.

The optimized structures of butane at different stretching points are shown in Fig.~\ref{fig:bu}. Increasing the distance between the terminal carbon atoms leads to a stretching of both external and internal C-C bonds. At a distance of 5.0 \AA, which corresponds to the maximum stretching force, the C-C bond lengths reach their maximum values. Further increasing the distance between the terminal carbon atoms leads to the formation of an ethylene molecule and two methyl radicals. Similarly, when the external force is applied gradually to the molecule and geometry relaxation is allowed during the stretching process, formation of the stable ethylene molecule favours the dissociation of the external C-C bonds.

\begin{figure}[t]
\includegraphics[scale=0.6]{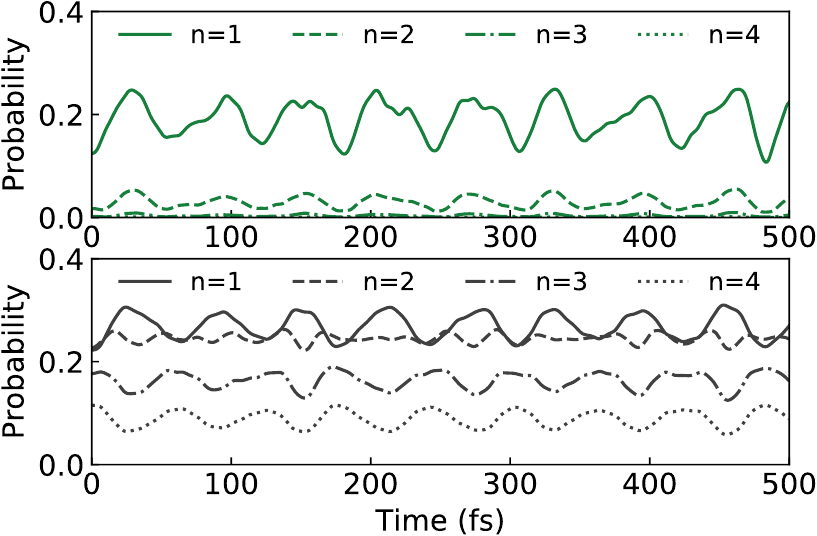}
\centering
\caption{Probability of observing different numbers of vibrational quanta ($n$) in the local modes of butane during a vibronic transition from ground to the force-modified potential energy surface plotted as a function of time. The local modes correspond to the stretching of the inner (top) and an outer (bottom) C-C bond. The magnitude of the external force is 3.88 nN.}\label{fig:budyn}
\end{figure}

The potential energy curve for stretching the external carbon atoms of butane is presented in Fig.~\ref{fig:buscanmarg}. When the external force is applied instantaneously to the terminal carbons of butane, the molecule is vertically transformed to a FM-PES and its vibrational modes get excited. We explore these excitations by investigating the marginal distributions of the vibrational quanta in two localized \cite{Jacob_2009} modes of butane that correspond to the stretching of the inner and outer C-C bonds after application of external forces with magnitudes up to 4.42 nN, which is about 0.75 of the critical force required to break the outer C-C bonds. The results are presented in Fig.~\ref{fig:buscanmarg}. When the magnitude of the external force is less than 3 nN, the vibrational excitations are very small, but the application of larger external forces lead to significant excitations in both inner and outer C-C bonds. More specifically, the localized mode that corresponds to the stretching of the inner C-C bond gets significantly excited when a force of 4.42 nN is applied (see Fig.~\ref{fig:buscanmarg}). This vibrational excitation helps with breaking the inner C-C bond which has a lower dissociation barriers at the FM-PES. This observation provides further evidence supporting the role of vibrational excitations in explaining the dissociation of the inner C-C bond in the sudden-force regime of butane mechanochemistry \cite{Rybkin_2017, Smalo_2014}.

\begin{table}[b]
\caption[]{Probabilities (in percentages) of simultaneously exciting outer and inner C-C stretching modes of butane, computed from ten thousand GBS samples obtained at different times t (in femtoseconds).}
\label{table:bucoex}
\begin{center}
\begin{tabular}{ c c c c c c c}
\hline
 Excitation\footnotemark & [1,1] & [1,2] & [2,1] & [1,3] & [2,2] & [1,4]  \\
 \hline
 t = 0   & 9.1 & 5.8 & 2.8 & 2.6 & 1.7 & 1.0\\
 t = 20  & 8.1 & 3.8 & 0.9 & 1.4 & 0.5 & 0.4\\
 t = 40  & 3.4 & 2.1 & 0.4 & 1.0 & 0.2 & 0.5\\ 
 t = 60  & 9.7 & 6.1 & 1.7 & 2.2 & 1.1 & 0.8\\
 t = 80  & 8.1 & 5.4 & 1.6 & 2.0 & 1.0 & 0.6\\
 t = 100 & 8.5 & 5.8 & 1.2 & 2.4 & 1.0 & 1.0\\ 
 \hline
\end{tabular}
\footnotetext{The number of vibrational quanta in the outer (n) and inner (m) C-C stretching modes are represented as [n,m].}
\end{center}
\end{table}

The distributions in Fig.~\ref{fig:buscanmarg} correspond to the instantaneous vibrational excitations during the transition to the FM-PESs. However, the distribution of the vibrational quanta in the local modes is time dependent and might fluctuate with time \cite{Sparrow_2018}. We look at the probability of the vibrational excitations in the localized modes of butane during the vibronic transition when a force of 3.88 nN is applied. For this value of the force, the vibrational energy levels of the inner C-C bond are not significantly populated when the vibrational dynamics is not included (see~Fig.~\ref{fig:buscanmarg}). In Fig.~\ref{fig:budyn}, the excitation probabilities for the C-C bonds are plotted with respect to time. The probability of exciting the inner C-C bond to its first vibrational energy level increases by about two times after only 25 femtoseconds. Analogous fluctuations are also observed for excitation to the higher energy levels of an outer C-C bond as illustrated in Fig.~\ref{fig:budyn}.

We now look at the co-excitation of the C-C stretching local modes of butane by generating GBS samples for the transition mediated by the external forces of 4.42 nN. In Table~\ref{table:bucoex}, the probability of vibrational co-excitation in the local modes of butane that correspond to the inner and outer C-C stretching are presented for different times. These data indicate that the co-excitation probability of these two modes is not significant, specially for higher excitation levels (e.g., up to only 1.7\% for double excitation in both modes). This is also in agreement with the observation that the simultaneous dissociation of both inner and outer C-C bond is not probable \cite{Rybkin_2017, Smalo_2014}. The sampling was performed for a reduced state of butane containing ten vibrational modes that were selected based on their contribution to the inner and outer C-C stretching local modes due to computational expenses of including all modes in the sampling.

\section{Conclusions}\label{sec:conc}

Recent advances in the development of quantum computers have opened the possibility to perform quantum simulation of materials on quantum devices. This creates a demand for developing quantum algorithms that can leverage the capabilities of near-term quantum computers for simulating molecular systems. In this work, we have demonstrated the ability of Gaussian boson sampling, which is a model for photonic quantum computing, to predict the vibrational excitation of molecules in vibronic processes. Predicting such excitations is important for understanding the impact of molecular vibrations on the outcome of chemical reactions that occur as a result of vibronic transitions.

We introduced and implemented an algorithm to simulate the vibrational excitation of pyrrole during a vibronic transition and also investigated the possibility of optimizing the number of vibrational quanta in specific vibrational modes by pre-exciting the vibrational modes of the ground electronic state. This application is motivated by experimental observations that demonstrated an enhancement in the dissociation of the N-H bond of pyrrole as a result of pre-excitation of the vibrational modes at the ground electronic state. Furthermore, we studied the excitation of vibrational modes in butane after application of mechanical forces. Such vibrational excitations can change the mechanism of bond dissociation in butane. Our results for these two model systems demonstrate that the vibrational excitations predicted by Gaussian boson sampling can help to explain experimental results and the predictions of molecular dynamics simulations.

Alongside the quantum algorithm, we introduced quantum-inspired classical methods that can be implemented in cases where only a small number of modes are of interest. The computational time required to perform sampling with these algorithms increases with the number of modes and using a Gaussian boson sampling device becomes necessary for large molecules and in cases where large numbers of modes need to be simulated.

We expect our results to motivate further investigations on the impact of controlled vibrational excitations in applications beyond light-induced vibronic transitions. An example of such applications is the change in the reactivity and electronic properties of molecules in single-molecule junctions.\\

\section*{Acknowledgments}
We thank Nathan Killoran and Christian Weedbrook for valuable discussions.\\

\noindent
\textbf{Conflicts of interest}\\
There are no conflicts of interest to declare.

\bibliographystyle{rsc}
\bibliography{MV_references}

\vspace{10pt}

%\newpage

\appendix

\section{Sampling probability in GBS}\label{app:gbs_prob}
In a GBS setting, the probability $\Pr(\bm{p})$ of observing an output $\bm{p}=(\bm{m},\bm{n}) = (m_1, m_2, \ldots, m_M,n_1, n_2, \ldots, n_M)$ is given by \cite{bjorklund2018faster,quesada2019franck,quesada2019simulating}:
\begin{align}\label{Eq: lhaf}
\Pr(\bm{p})  = \frac{\exp\left(-\tfrac{1}{2} \bm{\alpha}'^\dagger Q^{-1} \bm{\alpha}' \right)}{\prod_{i=1}^{2M} p_i!}  \frac{\text{lhaf}(\mathcal{A}'_{\bm{p}})}{\sqrt{\text{det}(Q)}},
\end{align}
where $\text{lhaf}(\cdot)$ is the loop hafnian defined as~\cite{bjorklund2018faster}:
\begin{align}\label{Eq: lhaf}
\text{lhaf}(A) = \sum_{\mu \in \text{SPM}(n)} \prod_{\scriptscriptstyle (i,j) \in \mu} A_{ij},
\end{align}
where SPM are the set of single-pair matchings, i.e., the perfect matchings of complete graph allowing for loops. Moreover, we define $Q:=V +\id/2$, $\bm{\alpha}' := (\bm{\alpha},\bm{\alpha}^*)^T$ and the matrix $\mathcal{A}'$ is given by
\beq
\mathcal{A}'_{\bm{p}} = \text{fdiag}(\mathcal{A}_{\bm{p}},\gamma_{\bm{p}}) 
%\mathcal{A}'_{ij} := \begin{cases}
%\mathcal{A}_{ij} &\text{ if } i\neq j,\\
%\gamma_{i} &\text{ if } i=j,
%\end{cases}
\eeq
where $\mathcal{A} := X \left(\id- Q^{-1}\right)$, $X :=  \left[\begin{smallmatrix}
	0 &  \id \\
	\id & 0  
\end{smallmatrix} \right]$, and $\gamma_{i}$ is the $i^\text{th}$ entry of $\gamma := Q^{-1} \bm{\alpha}'$ and $\text{fdiag}$ is the function that places in the diagonal of its first argument the entries of its second argument. The matrix $\mathcal{A}_{\bm{p}}$ is obtained by taking the original matrix $\mathcal{A}$ and repeating rows and columns $i$ and $i+m$ a total of $p_i$ times. If $p_i=0$ then rows and columns $i$ and $i+M$ are deleted.

\section{Initial state preparation}\label{app:normal_modes}
We derive Eq.~\eqref{Eq:time-evolution} in the main text. Recall the expression for the time-evolution operator
\begin{align}
\hat{\mathcal{U}}(t_0,t_1) = \hat{\mathcal{T}}\exp\left( - \frac{i}{\hbar} \int_{t_0}^{t_1} dt \hat{H}_{\text{LM}}^{I}(t)  \right).
\end{align}
The light Hamiltonian does not commute with itself at different times, yet the only effect of time ordering is to add an overall global phase to the expression obtained by removing $\hat{\mathcal{T}}$ in the last expression (see Sec. 2.1.3 of Ref.~\cite{quesada2015very}). We can now integrate the Hamiltonian to obtain
\begin{widetext}
\begin{align}
\int_{t_0}^{t_1} dt \hat{H}_{\text{LM}}^{I}(t) =& \int_{t_0}^{t_1} dt \sum_i q_i \hat{C}_i \cdot \left( \bm{E}_0 e^{-i \omega_0 t} +\bm{E}_0^* e^{i \omega_0 t} \right) \\
=&\sum_i q_i \sum_{j} \left( \bm{c}_{ij}\cdot \bm{E_0} \hat{a}_j^\dagger \left( \frac{e^{i(\omega_j - \omega_0)t_1}- e^{i(\omega_j - \omega_0)t_0} }{i(\omega_j - \omega_0)} + \frac{e^{i(\omega_j + \omega_0)t_1}- e^{i(\omega_j + \omega_0)t_0} }{i(\omega_j + \omega_0)} \right) + \text{H.c.} \right).
\end{align}
\end{widetext}
Now assume that one of the normal modes, $j=k$ is nearly resonant with the radiation $\omega_j \approx \omega_0$, which implies that this term will dominate over any other term in the last equation. Noticing the following limit,
\begin{align}
\lim_{\omega_k \to \omega_0}\frac{e^{i(\omega_k - \omega_0)t_1}- e^{i(\omega_k - \omega_0)t_0} }{i(\omega_k - \omega_0)} = t_1-t_0,
\end{align}
we can write at resonance with mode $k$, 
\begin{align}\label{Eq:time-evolution_b}
\hat{\mathcal{U}}(t_0,t_1) &= \exp\left(- \frac{i}{\hbar} \sum_i q_i  \bm{c}_{ik}\cdot \bm{E_0} \hat{a}_k^\dagger (t_1-t_0) - \text{H.c.} \right)\\
&= \hat{D}_k\left(- \frac{i}{\hbar} \sum_i q_i  \bm{c}_{ik}\cdot \bm{E_0} \  \{t_1-t_0\} \right).
\end{align}

\end{document}